\documentclass{article}
\usepackage[preprint]{spconf}
\usepackage{amsmath,graphicx,url,times,amsmath,amssymb,acronym,balance}
\usepackage{color,booktabs}
\usepackage[table]{xcolor}
\usepackage{cite}
\usepackage{lipsum}

\copyrightnotice{\copyright\ IEEE 2021}
 \toappear{Submitted to {\it Proc.\ ICASSP2022, May 2022, Singapore}}

\title{Effect of noise suppression losses\\ on speech distortion and ASR performance}
\name{Sebastian Braun, Hannes Gamper} 
\address{
	Microsoft Research, Redmond, WA, USA \\
	\{sebastian.braun, hannes.gamper\}@microsoft.com}
\acrodef{STFT}{short-time Fourier transform}
\acrodef{MSE}{mean-squared error}
\acrodef{MAE}{mean absolute error}
\acrodef{PSD}{power spectral density}
\acrodef{RTF}{relative transfer function}
\acrodef{SNR}{signal-to-noise ratio}
\acrodef{segSNR}{segmental signal-to-noise ratio}
\acrodef{SRR}{signal-to-reverberation ratio}
\acrodef{PDF}{probability density function}
\acrodef{DOA}{direction-of-arrival}
\acrodef{VAD}{voice activity detector}
\acrodef{MVDR}{minimum variance distortionless response}
\acrodef{AIR}{acoustic impulse response}
\acrodef{PESQ}{perceptual evaluation of speech quality}
\acrodef{STOI}{short-time objective intelligibility}
\acrodef{LSD}{log spectral distance}
\acrodef{CD}{cepstral distance}
\acrodef{WER}{word error rate}
\acrodef{SPP}{speech presence probability}
\acrodef{DNN}{deep neural network}
\acrodef{RNN}{recurrent neural network}
\acrodef{CNN}{convolutional neural network}
\acrodef{FC}{fully connected}
\acrodef{CRN}{convolutional recurrent network}
\acrodef{LSTM}{long-term short-term}
\acrodef{GRU}{gated recurrent unit}
\acrodef{FF}{feed forward}
\acrodef{ReLU}{rectified linear unit}
\acrodef{GCC}{generalized cross-correlation}
\acrodef{RMSE}{root-mean-square error}
\acrodef{CPSD}{cross-power spectral density}
\acrodef{siSDR}{scale-invariant signal-to-distortion ratio}
\acrodef{SDR}{signal-to-distortion ratio}
\acrodef{MagMSE}{Magnitude MSE}
\acrodef{LPS}{logarithmic power spectrum}
\acrodef{CSE}{complex spectrum error}
\acrodef{MagSE}{magnitude spectrum error}
\acrodef{LogMagSE}{logarithmic magnitude spectrum error}
\acrodef{DL}{deep learning}
\acrodef{PLSD}{phase-aware logarithmic spectrum distance }
\acrodef{SDW}{speech distortion-weighted}
\acrodef{MOS}{mean opinion score}
\acrodef{RIR}{room impulse response}
\acrodef{MAC}{multiply-accumulate}
\acrodef{CRUSE}{Convolutional Recurrent U-net for Speech Enhancement}
\acrodef{FD}{frequency-domain}
\acrodef{DNS}{deep noise suppression}
\acrodef{SED}{sound event detection}
\acrodef{ASR}{automatic speech recognition}
\acrodef{ERB}{equivalent rectangular bandwidth}
\acrodef{RSNR}{reverberant speech-to-noise ratio}

\newcommand{\F}[2]{\mathcal{F}_\text{#1}\left\{#2\right\}}

\newcommand{\seb}[1]{{\color{blue}#1}}

\definecolor{matlab1}{rgb}{0, 0.4470, 0.7410}
\definecolor{matlab2}{rgb}{0.8500, 0.3250, 0.0980} 
\definecolor{matlab3}{rgb}{0.9290, 0.6940, 0.1250} 
\definecolor{matlab4}{rgb}{0.4940, 0.1840, 0.5560} 
\definecolor{matlab5}{rgb}{0.4660, 0.6740, 0.1880}

\begin{document}
	\ninept
	
	\maketitle
	\begin{abstract}
		Deep learning based speech enhancement has made rapid development towards improving quality, while models are becoming more compact and usable for real-time on-the-edge inference. However, the speech quality scales directly with the model size, and small models are often still unable to achieve sufficient quality. Furthermore, the introduced speech distortion and artifacts greatly harm speech quality and intelligibility, and often significantly degrade automatic speech recognition (ASR) rates. 
		In this work, we shed light on the success of the spectral complex compressed mean squared error (MSE) loss, and how its magnitude and phase-aware terms are related to the speech distortion vs.\ noise reduction trade off. We further investigate integrating pre-trained reference-less predictors for mean opinion score (MOS) and word error rate (WER), and pre-trained embeddings on ASR and sound event detection. Our analyses reveal that none of the pre-trained networks added significant performance over the strong spectral loss.
	\end{abstract}
	

	\section{Introduction}
	\label{sec:intro}
	Speech enhancement techniques are present in almost any device with voice communication or voice command capabilities. The goal is to extract the speaker's voice only, reducing disturbing background noise to improve listening comfort, and aid intelligibility for human or machine listeners.
	In the past few years, neural network-based speech enhancement techniques showed tremendous improvements in terms of noise reduction capability~\cite{Wang2018,Reddy2021a}. Data-driven methods can learn the tempo-spectral properties of any type of speech and noise, in contrast to traditional statistical model-based approaches that often mismatch certain types of signals.
	While one big current challenge in this field is still to find smaller and more efficient network architectures that are computationally light for real-time processing while delivering good results, another major challenge addressed here is obtaining good, natural sounding speech quality without processing artifacts.
	
	In the third \ac{DNS} challenge \cite{Reddy2021a}, the separate evaluation of speech distortion (SIG), background noise reduction (BAK), and overall quality (OVL) according to ITU P.835 revealed that while current state-of-the-art methods achieve outstanding noise reduction, only one submission did not degrade SIG on average while still showing high BAK. Degradations in SIG potentially also harm the performance of following \ac{ASR} systems, or human intelligibility as well.
	
	In \cite{Eskimez2021} a speech enhancement model was trained on a signal-based loss and a \ac{ASR} loss with alternating updates. This method requires either transcriptions for all training data, or using a transcribed subset for the \ac{ASR} loss. The authors also proposed to update the \ac{ASR} model during training, which creates an in practice often undesired dependency between speech enhancement and the jointly trained \ac{ASR} engine. 
	
	In this work, we explore several loss functions for a real-time deep noise suppressor with the goal to improve SIG without harming \ac{ASR} rates. The contribution of this paper is threefold.
	First, we show that by decoupling the loss from the speech enhancement inference engine using end-to-end training, choosing a higher resolution in a spectral signal-based loss can improve SIG. Second, we propose ways to integrate MOS and WER estimates from pre-trained networks in the loss as weighting. Third, we evaluate additional supervised loss terms computed using pre-trained networks, similar to the deep feature loss \cite{Germain2018}. In \cite{Kataria2021}, six different networks pre-trained on different tasks have been used to extract embeddings from output and target signals, to form an additional loss, where a benefit was reported only for the sound event detection model published in \cite{Kong2020}. We show different results when trained on a larger dataset and evaluated on real data using decisive metrics on speech quality and \ac{ASR} performance, while we find that none of the pre-trained networks improved \ac{ASR} rates or speech quality significantly.

	\section{System and training objective}
	\label{sec:enhancement_system}
	A captured microphone signal can generally be described by
	\begin{equation}
		y(t) = m\left\{ s(t) + r(t) + v(t) \right\},
	\end{equation}
	where $s(t)$ is the non-reverberant desired speech signal, $r(t)$ undesired late reverberation, $v(t)$ additive noise or interfering sounds, and $m\{\}$ can model linear and non-linear acoustical, electrical, or digital effects that the signals encounter.
	
	\subsection{End-to-end optimization}
\begin{figure}
		\centering
		\includegraphics[width=\columnwidth,clip,trim=180 100 180 90]{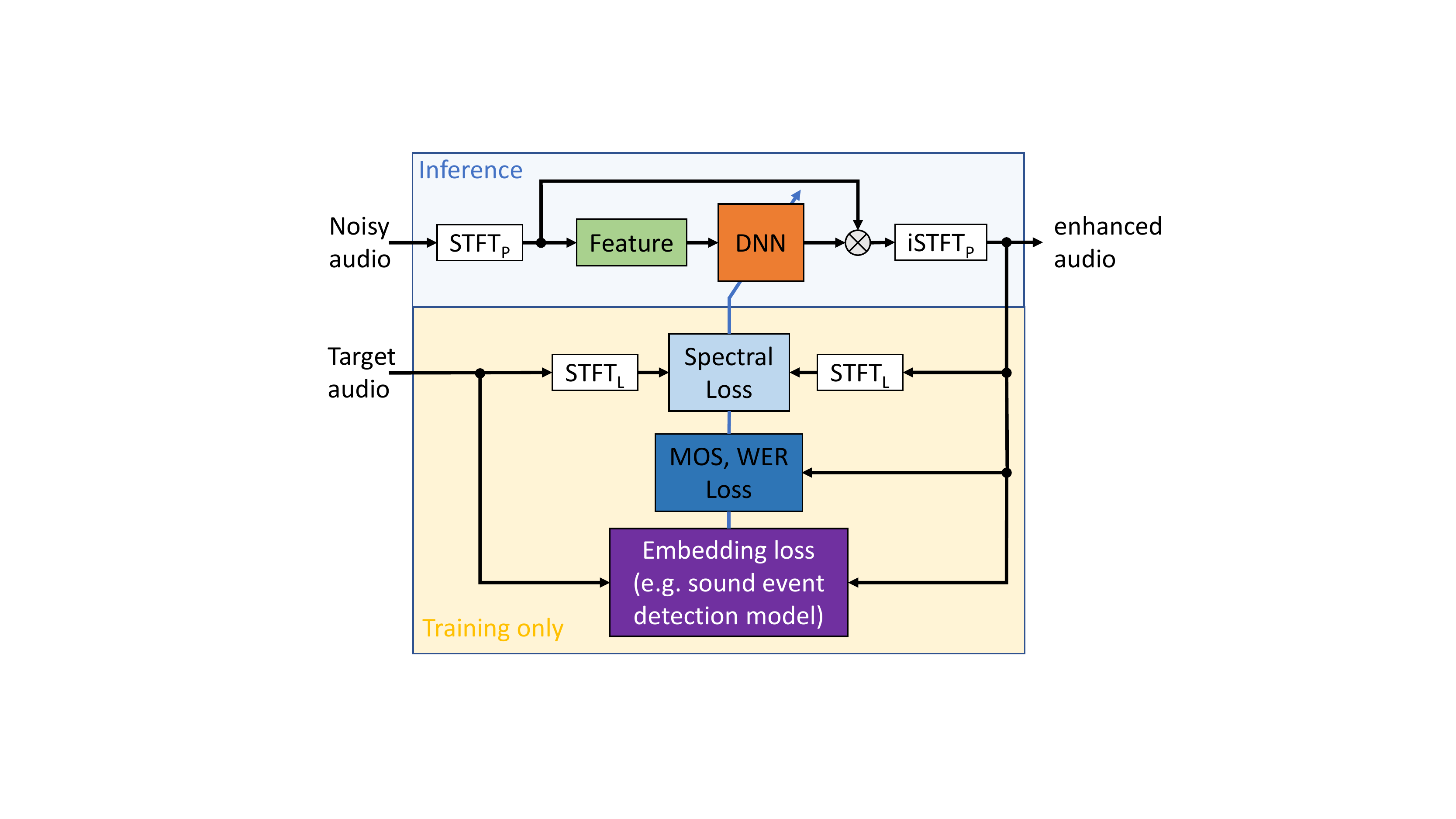}
		\caption{End-to-end trained system on various losses.}
		\label{fig:end2end}
	\end{figure}

	We use an end-to-end enhancement system using a complex enhancement filter in the \ac{STFT} domain
	\begin{equation}
		\label{eq:shat}
		\widehat{s}(t) = \F{P}{ G(k,n) \; \F{P}{y(t)} }^{-1}
	\end{equation}
	where $\F{P}{a(t)} = A(k,n)$ denotes the linear \ac{STFT} operator yielding a complex signal representation at frequency $k$ and time index $n$, and $G(k,n)$ is a complex enhancement filter. The end-to-end optimization objective is to train a neural network predicting $G(k,n)$, while optimizing a loss on the time-domain output signal $\widehat{s}(t)$ as shown in Fig.~\ref{fig:end2end}.
	
	\subsection{Features and network architecture}
	We use complex compressed features by feeding the real and imaginary part of the complex FFT spectrum as channels into the first convolutional layer. Magnitude compression is applied to the complex spectra by
	\begin{equation}
		Y^c(k,n) = |Y(k,n)|^c \, \frac{Y(k,n)}{\max(|Y(k,n)|, \eta)},
	\end{equation}
	where the small positive constant $\eta$ avoids division by zero.
	
	We use the \ac{CRUSE} model proposed in \cite{Braun2021a} with 4 convolutional encoder/decoder layers with time-frequency kernels (2,3) and stride (1,2) with pReLU activations, a group of 4 parallel GRUs in the bottleneck and add skip connections with $1\times1$ convolutions.
	The network output are two channels for the real and imaginary part of the complex filter $G(k,n)$. To ensure stability, we use \emph{tanh} output activation functions restraining the filter values to $[-1,1]$ as in \cite{Choi2019}.

	\section{Data generation and augmentation}
	We use an online data generation and augmentation technique, using the power of randomness to generate virtually infinitely large training data. Speech and noise portions are randomly selected from raw audio files with random start times to form 10~s clips. If a section is too short, one or more files are concatenated to obtain the 10~s length. 80\% of speech and noise are augmented with random biquad filters \cite{Valin2018} and 20\% are pitch shifted within $[-2,8]$ semi-tones. If the speech is non-reverberant, a random \ac{RIR} is applied. The non-reverberant speech training target is obtained by windowing the \ac{RIR} with a cosine decay of length 50~ms, starting 20~ms (one frame) after the direct path. Speech and noise are mixed with a \ac{SNR} drawn from a normal distribution with mean and standard deviation $\mathcal{N}(5, 10)$~dB. The signal levels are varied with a normal distribution $\mathcal{N}(-26, 10)$~dB.
	
	We use 246~h noise data, consisting of the \ac{DNS} challenge noise data (180~h), internal recordings (65~h), and stationary noise (1~h), as well as 115~k \acp{RIR} published as part of the \ac{DNS} challenges \cite{Reddy2021}. Speech data is taken mainly from the 500~h high quality-rated Librivox data from \cite{Reddy2021}, in addition to high \ac{SNR} data from AVspeech \cite{Ephrat2018} and the Mandarin and Spanish, singing, and emotional CremaD corpora published within \cite{Reddy2021}, an internal collection of 8~h emotional speech and 2~h laughing sourced from Freesound. 
	\begin{figure}
		\centering
		\includegraphics[width=0.95\columnwidth,clip,trim=20 0 30 15]{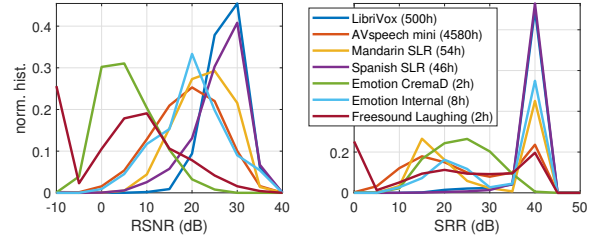}
		\caption{SNR and SRR distributions of speech datasets}
		\label{fig:speech_snr}
	\end{figure}
	Figure~\ref{fig:speech_snr} shows the distributions of \ac{RSNR} and \ac{SRR} as predicted by a slightly modified neural network following \cite{Braun2021}. The singing data is not shown in Fig.~\ref{fig:speech_snr} as it is studio quality and our RSNR estimator is not trained on singing. While the Librivox data has both high RSNR and SRR, this is not the case for the other datasets, which have broader distributions and lower peaks. Therefore, we select only speech data from the AVspeech, Spanish, Mandarin, emotion, and laughing subsets with $segSNR > 30$~dB and $SRR>35$~dB for training.

	\section{Loss functions}
	\label{sec:losses}
	In this section, we describe training loss functions that are used to optimize the enhanced signal $\widehat{s}(t)$. We always use a standard signal-based spectral loss described in Sec.~\ref{sec:ccmse}, which is extended or modified in several ways as described in the following subsections.
	
	\subsection{Magnitude-regularized compressed spectral loss}
	\label{sec:ccmse}
	As a standard spectral distance-based loss function $\mathcal{L}_{SD}$, we use the complex compressed loss \cite{Ephrat2018,Lee2018}, which outperformed other spectral distance-based losses in \cite{Braun2021b}, given by
	\begin{equation}
		\label{eq:loss}
		\mathcal{L}_{SD} = \frac{1}{\sigma_s^c}\left( \lambda \sum_{\kappa,\eta} \left|S^c \!-\! \widehat{S}^c \right|^2 + (1\!-\!\lambda) \sum_{\kappa,\eta} \left||S|^c \!-\! |\widehat{S}|^c\right|^2 \right),
	\end{equation}
	where here the spectra $S(\kappa,\eta) = \F{L}{s(t)}$ and $\widehat{S}(\kappa,\eta) = \F{L}{\widehat{s}(t)}$ are computed with a \ac{STFT} operation with independent settings from $\F{P}{\cdot}$ in \eqref{eq:shat}, $A^c = |A|^c \frac{A}{|A|}$ is a magnitude compression operation, and the frequency and time indices $\kappa,\eta$ are omitted for brevity. The loss for each sequence is normalized by the active speech energy $\sigma_s$ \cite{Braun2020}, which is computed from $s(t)$ using a \ac{VAD}. The complex and magnitude loss terms are linearly weighted by $\lambda$, and the compression factor is $c=0.3$.
	The processing \ac{STFT} $\mathcal{F}_\text{P}$ and loss frequency transform $\mathcal{F}_\text{L}$ can differ in e.\,g.\ window and hop-size parameters, or be even different types of frequency transforms as shown in Fig.~\ref{fig:end2end}.
	In the following subsections, we propose several extensions to the spectral distance loss $\mathcal{L}_{SD}$ to potentially improve quality and generalization.
	
	\textbf{Optional frequency weighting}
	Frequency weightings for simple spectral distances are often used in perceptually motivated evaluation metrics 
	and have been tried to integrate as optimization targets for speech enhancement \cite{Zhao2019}. While we already showed that the AMR-wideband based frequency weighting did not yield improvements for the experiments in \cite{Braun2021b}, here we explore another attempt applying a simple \ac{ERB} weighting \cite{Moore1983} to the spectra in \eqref{eq:loss} using 20 bands.
	
	\subsection{Additional cepstral distance term}
	The \ac{CD} \cite{Kitawaki1988} is one of the few intrusive objective metrics that is also sensitive to speech distortion artifacts caused by speech enhancement algorithms, in contrast to most other metrics, which are majorly sensitive to noise reduction. This motivated adding a \ac{CD} term $\mathcal{L}_{CD}$ to \eqref{eq:loss} by
	\begin{equation}
		\label{eq:cd_loss}
		\mathcal{L}_{CD} = \beta \mathcal{L}_{SD}(s, \widehat{s}) + (1-\beta) \mathcal{L}_{CD}(s, \widehat{s})
	\end{equation}
	where we chose $\beta=0.001$, but did not find a different weight that helped improving the results.
	
	\subsection{Non-intrusive speech quality and ASR weighted loss}
	Secondly, we explore the use of non-intrusive estimators for \ac{MOS} \cite{Gamper2019} and \ac{WER} \cite{Gamper2020a}. The \ac{MOS} predictor has been re-trained with subjective ratings of various speech enhancement algorithms, including the MOS ratings included in the three \ac{DNS} challenges. Note that both are blind (non-intrusive) estimators, meaning they give predictions without requiring any reference, which makes them also interesting for unsupervised training, to be explored in future work.
	To avoid dependence of a hyper-parameter when extending the loss function by additive terms (e.g., $\lambda$ in \eqref{eq:loss}), we use the predictions to weight the spectral distance loss for each sequence $b$ in a training batch by
	\begin{equation}
		\mathcal{L}_{MOS,WER} = \sum_b \frac{nWER(\widehat{s}_b)}{nMOS(\widehat{s}_b)} \, \mathcal{L}_{SD}(s_b, \widehat{s}_b),
	\end{equation}
	where $s_b(t)$ is the $b$-th sequence, $nWER()$ and $nMOS()$ are the \ac{WER} and \ac{MOS} predictors. We also explore MOS and WER only weighted losses by setting the corresponding other prediction to one.

	\subsection{Multi-task embedding losses}
	As a third extension, similiar to \cite{Kataria2021}, we add a distance loss  using pre-trained embeddings on different audio tasks, such as \ac{ASR} or \ac{SED}, by 
	\begin{equation}
		\mathcal{L}_\text{emb} = \sum_b \mathcal{L}_{SD}(s_b, \widehat{s}_b) + \gamma \frac{\Vert\mathbf{u}(s_b) - \mathbf{u}(\widehat{s}_b) \Vert_p}{\Vert\mathbf{u}(s_b)\Vert_p}
	\end{equation}
	where $\mathbf{u}(s_b)$ and $\mathbf{u}(\widehat{s}_b)$ are the embedding vectors from the target speech and output signals, respectively, and we use the normalized $p$-norm as distance metric.
	This differs from \cite{Kataria2021}, where a L\textsubscript{1} spectral loss was used for $\mathcal{L}_{SD}$. We verified a small benefit from the normalization term and chose $p$ according to the embedding distributions in preliminary experiments.
	
	In this work, we use two different embedding extractors $\mathbf{u}()$: a) the \emph{PANN} \ac{SED} model \cite{Kong2020} that was the only embedding that showed a benefit in \cite{Kataria2021}, and b) an \ac{ASR} embedding using \emph{wav2vec~2.0} models \cite{Baevski2020}.
	For \emph{PANN}, we use the pre-trained 14-layer CNN model using the first 2-6 double CNN layers with $p\!=\!1$ \seb{and $\gamma\!=\!0.05$}.
	For \emph{wav2vec~2.0}, we explore three versions of pre-trained models with $p\!=\!2$ \seb{and $\gamma\!=\!0.1$} to extract embeddings, which could help to improve \ac{ASR} performance of the speech enhancement networks: i) the small \emph{wav2vec-base} model trained on LibriSpeech, ii) the large \emph{wav2vec-lv60} model trained on LibriSpeech, and iii) the large \emph{wav2vec-robust} model trained on Libri-Light, CommonVoice, Switchboard, Fisher, i.\,e.\,, more realistic and noisy data. We use the full wav2vec models taking the logits as output. The pre-trained embeddings extractor networks are frozen while training the speech enhancement network.
	
	
	\begin{figure}[tb]
		\centering
		\includegraphics[width=0.9\columnwidth,clip,trim=0 0 10 5]{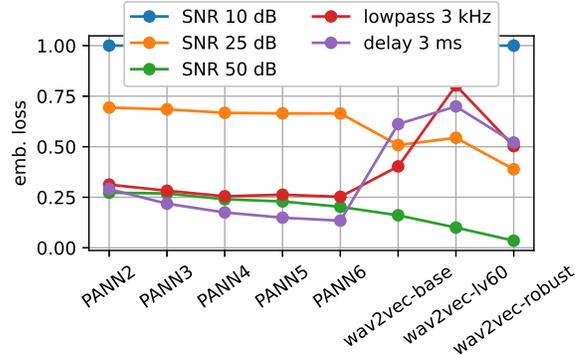}
		\vspace{-.3cm}
		\caption{Sensitivity of embedding losses to various degradations. Losses are normalized per embedding type (column).}
		\label{fig:emb-loss-sens}
	\end{figure}
	\textbf{Understanding the embeddings}
	To provide insights into how to choose useful embedding extractors and reasons why some embeddings work better than others, we conduct a preliminary experiment. 
	We show the embedding loss terms for a selection of signal degradations, corrupting a speech signal with the same noise with three different \acp{SNR}, a 3~kHz lowpass degradation, and the impact of a delay, i.\,e.\, a linear phase shift. The degradations are applied on 20 speech signals with different noise signals and results are averaged. Fig.~\ref{fig:emb-loss-sens} shows the PANN embedding loss using a different number of CNN layers, and the three wav2vec models mentioned above. The embedding loss is normalized to the maximum per embedding (i.e., per column) for better visibility, as we are interested in the differences created by certain degradations. 
	
	We observe that the PANN models are rather non-sensitive to lowpass degradations, attributing it a similar penalty as a -50~dB background noise. The wav2vec embeddings are much more sensitive to the lowpass distortion, and rate moderate background noise with -25~dB comparably less harmful than the PANN embeddings. This might be closer to a human importance or intelligibility rating, where moderate background noise might be perceived less disturbing than speech signal distortions, and seems therefore more useful in guiding the networks towards preserving more speech components rather than suppressing more (probably hardly audible) background noise.
	We consequently choose the 4-layer PANN and \emph{wav2vec-robust} embeddings for our later experiments.

	\section{Evaluation}
	\subsection{Test sets and metrics}
	\label{sec:testsets}
	We show results on the public third \ac{DNS} challenge test set consisting of 600 actual device recordings under noisy conditions \cite{Reddy2021a}. The impact on subsequent \ac{ASR} systems is measured using three subsets: The transcribed DNS challenge dataset (challenging noisy conditions), a collection of internal real meetings (18h, realistic medium noisy condition), and 200 high quality high SNR recordings.
	
	Speech quality is measured using a non-intrusive P.835 DNN-based estimator similar to DNSMOS \cite{Reddy2021b} trained on the available data from the three \ac{DNS} challenges and internally collected data. 
	We term the non-intrusive predictions for \emph{signal distortion}, \emph{background noise} and \emph{overall quality} from P.835 DNSMOS \emph{nSIG, nBAK, nOVL}. The P.835 DNSMOS model predicts SIG with $>0.9$ correlation and BAK, OVL with $>0.95$ correlation per model. 
	The impact on production-grade \ac{ASR} systems is measured using the public Microsoft Azure Speech SDK service \cite{azurespeechsdk_webpage} for transcription. 
	
	\subsection{Experimenal settings}
	The CRUSE processing parameters are implemented using a \ac{STFT} with square-root Hann windows of 20~ms length, 50\% overlap, and a FFT size of 320. 
	To achieve results in line with prior work,
	we use a network size that is on the larger side for most CPU-bound real-world applications, although it still runs in real-time on standard CPUs and is several times less complex than most research speech enhancement architectures:
	The network has 4 encoder conv layers with channels $[32,64,128,256]$, which are mirrored in the decoder, a GRU bottleneck split in 4 parallel subgroups, and conv skip connections \cite{Braun2021a}. The resulting network has 8.4~M trainable parameters, 12.8~M MACs/frame and the ONNX runtime has a processing time of 45~ms per second of audio on a standard laptop CPU. For reference, the first two ranks in the 3rd DNS challenge \cite{Reddy2021a} stated to consume about 60~M MACs and 93~M FLOPs per frame.
	The network is trained using the AdamW optimizer with initial learning rate 0.001, which is halved after plateauing on the validation metric for 200 epochs. Training is stopped after validation metric plateau of 400 epochs. One epoch is defined as 5000 training sequences of 10~s. We use the synthetic validation set and heuristic validation metric proposed in \cite{Braun2021a}, a weighted sum of PESQ, siSDR and \ac{CD}.
	
	\subsection{Results}

	\begin{figure}[tb]
		\centering
		\includegraphics[width=.85\columnwidth,clip,trim=0 0 10 5]{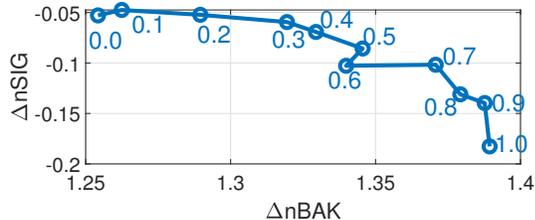}
		\vspace{-.2cm}
		\caption{Controlling the speech distortion - noise reduction tradeoff for \eqref{eq:loss} using the  {\color{matlab1}complex loss weight $\lambda$}, where $\lambda=1$ gives the complex loss term only, and $\lambda=0$ gives the magnitude only loss.}
		\label{fig:ccomp-weight}
	\end{figure}
	In the first experiment, we study the not well understood linear weighting between complex compressed and magnitude compressed loss \eqref{eq:loss}. Fig.~\ref{fig:ccomp-weight} shows the nSIG vs.\ nBAK tradeoff when changing the contribution of the complex loss term with the weight $\lambda$ in \eqref{eq:loss}. The magnitude term acts as a regularizer for distortion, while higher weight on the complex term yields stronger noise suppression, at the price of speech distortion. An optimal weighting is therefore found as a trade off. We use $\lambda=0.3$ in the following experiments, which we found to be a good balance between SIG and BAK.
	
	Table~\ref{tab:loss_results} shows the results in terms of nSIG, nBAK, nOVL and WER for all losses under test, where \emph{DNS}, \emph{HQ}, and \emph{meet} refer to the three test sets described in Sec.~\ref{sec:testsets}.
	\begin{table}[tb]
		\addtolength{\tabcolsep}{-2pt}  
		\centering
		\begin{tabular}{l|ccc|ccc}
			\toprule
			loss 	& nSIG & nBAK & nOVL &\multicolumn{3}{c}{WER (\%)} \\
			\hspace{1.3cm} dataset & \multicolumn{3}{c|}{DNS} & DNS & HQ & meet \\
			\midrule
			noisy 			& 3.87	& 3.05	& 3.11 & 27.9	& 5.7	& 16.5 \\
			$\mathcal{L}_{SD}^{20}$ (20~ms, 50\%) 	& 3.77	& 4.23	& 3.50 & 31.0	& 5.9	& 18.7 \\
			$\mathcal{L}_{SD}^{32}$ (32~ms, 50\%) 	& 3.79	& 4.26	& 3.53 & 30.6	& 5.9	& 18.6 \\ 
			$\mathcal{L}_{SD}^{64}$ (64~ms, 75\%) 	& \textbf{3.79}	& \textbf{4.28}	& \textbf{3.54} & \textbf{30.1}	& 5.9	& 18.4 \\
			$\mathcal{L}_{SD}^{64}$ ERB		& 3.73	& 4.22	& 3.46 & 31.9	& 6.0	& 18.6 \\
			$\mathcal{L}_{SD}^{64}$ + CD	& \textbf{3.79}	& 4.26	& \textbf{3.53}	& 30.4	& \textbf{5.8}	& \textbf{18.1} \\
			$\mathcal{L}_{SD}^{64}$-MOS 		& 3.78	& \textbf{4.27}	& \textbf{3.53} & 30.2	& 6.0	& \textbf{18.0} \\
			$\mathcal{L}_{SD}^{64}$-WER 		& \textbf{3.79}	& \textbf{4.27}	& \textbf{3.53} & 30.5	& \textbf{5.8}	& 18.2 \\
			$\mathcal{L}_{SD}^{64}$-MOS-WER 	& \textbf{3.79}	& 4.26	& \textbf{3.53} & \textbf{30.1}	& \textbf{5.8}	& 18.4 \\
			$\mathcal{L}_{SD}^{64}$ + PANN\textsubscript{4}		& \textbf{3.79}	& \textbf{4.27}	& \textbf{3.54} & 30.4 	& \textbf{5.8}	& 18.5 \\
			$\mathcal{L}_{SD}^{64}$ + wav2vec 	& \textbf{3.79}	& 4.26	& \textbf{3.53} & 30.3	& \textbf{5.9}	& 18.6 \\
			\bottomrule
		\end{tabular}
		\caption{Impact of modifying and extending the spectral loss on perceived quality and ASR.}
		\label{tab:loss_results}
		\addtolength{\tabcolsep}{2pt}  
	\end{table}
	The first three rows show the influence of the \ac{STFT} resolution $\F{L}{\cdot}$ used to compute the end-to-end complex compressed loss \eqref{eq:loss}, where we used Hann windows of $\{20,32,64\}$~ms with $\{50\%,50\%,75\%\}$ overlap. The superscript of $\mathcal{L}_{SD}$ indicates the STFT window length. We can observe that the larger windows lead to improvements of all metrics. This is an interesting finding, also highlighting the fact that speech enhancement approaches implemented on window sizes or look-ahead are not comparable. With the decoupled end-to-end training, we can improve performance with larger \ac{STFT} resolutions of the loss, while keeping the smaller \ac{STFT} resolution for processing, to keep the processing delay low.
	Similar to many other frequency weightings, the \ac{ERB}-weighted spectral distance did not work well, showing a significant degradation compared to the linear frequency resolution.
	
	Contrary to expectations, the additive \ac{CD} term did not help to improve SIG further, but slightly reduced BAK. It did however improve the WER for the high quality and meeting test data.
	Disappointingly, the non-intrusive MOS weighting did not improve any MOS metrics over the plain best spectral distance loss, and shows no clear trend for WER. A reason could be that overall MOS is still emphasizing BAK more, whereas we would need a loss that improves SIG to achieve a better overall result. The non-intrusive WER weighting shows a minor WER improvement for the high-quality and meeting data, with small degradation of DNS testset compared to $\mathcal{L}_{SD}^{64}$ only. As the ASR models to train the non-intrusive WER predictor were trained on mostly clean speech, this could be a reason for the WER model not helping the noisy cases. 
	The MOS+WER weighting ranks in between MOS and WER only weighted losses.
	
	Tab.~\ref{tab:loss_results} shows results for the PANN-augmented loss using the 4-layer PANN only, which also does not show an improvement. Using more \emph{PANN} layers or a higher PANN weight $\gamma>0.05$ resulted in worse performance, and could not exceed the standalone $\mathcal{L}_{SD}$ loss. Possible  reasons have already been indicated in Fig.~\ref{fig:emb-loss-sens}. The \emph{wav2vec} ASR embedding loss term shows also no significant improvement in terms of WER or MOS.
	Note that the P.835 DNSMOS absolute values, especially $nOVL$ are somewhat compressed. The CRUSE model with $\mathcal{L}_{SD}^{20}$ achieved $\Delta SIG=-0.17$, $\Delta BAK=1.98$, $\Delta OVL=0.85$ in the subjective P.835 tests \cite{DNSCO2021}.

	\section{Conclusions}
	In this work, we provided insight into the advantage of magnitude regularization in the complex compressed spectral loss to trade off speech distortion and noise reduction. We further showed that an increased spectral resolution of this loss can lead to significantly better results. Besides the increased resolution loss, modifications that also aimed at improving signal quality and distortion, e.\,g.\, integrating pre-trained networks could not provide a measurable improvement. Also, loss extensions that introduced knowledge from pre-trained \ac{ASR} systems showed no improvements in generalization for human or machine listeners.
	The small improvements in signal distortion and WER indicate that there is more research required to improve these metrics significantly.

	\vfill\pagebreak
	\balance
	
	\bibliographystyle{IEEEbib}
	\bibliography{sapref.bib,localbib.bib}
	
\end{document}